\documentclass[twocolumn,showpacs,preprintnumbers,amsmath,amssymb, floatfix]{revtex4}

\usepackage{graphicx}
\usepackage{dcolumn}
\usepackage{bm}

\newcommand{\Eq}[1]{Eq.~(\ref{#1})}

\begin{document}

\title{A Renormalization Group Analysis of Coupled Superconducting and Stripe Order in 1+1 Dimensions}
\author{Henry C. Fu}
\affiliation{Department of Physics,University of California at Berkeley,
Berkeley, CA 94720, USA}

\date{\today}

\begin{abstract}
In this paper we perform a renormalization group analysis on the 1+1
dimensional version of a previously proposed effective field theory
\cite{Lee} describing (quantum) fluctuating stripe and superconductor
orders. We find four possible phases corresponding to stripe
order/disorder combined with superconducting order/disorder.
\end{abstract}

 \maketitle

\section{Introduction}
In $La_{2-x}Sr_xCuO_4$ (LSCO) compounds, there are three well-established
ordering tendencies: antiferromagnetism, superconductivity and charge/spin
stripes.\cite{Tranquada}. Some experiments indicate that stripes and
superconductivity  can even coexist in these compounds\cite{coexist}.
Furthermore, neutron scattering experiments by Lake {\it et
al.}\cite{Lake} show that a moderate magnetic field can have large effects
on the incommensurate magnetic fluctuations. This is widely taken as
evidence suggesting the stabilization of stripes by the magnetic field.

In mean field theory, when two order parameters are in close competition
it is possible for them to coexist in a certain region of the phase
diagram\cite{Kivelson}. In such a coexistence region quantum fluctuations
of both order parameters dominate the low-energy physics. In a recent
paper Lee\cite{Lee} examined such a situation.  That paper described how
the Goldstone modes of stripe and superconducting orders and their
respective topological defects interact.

We stress that the theory presented in Ref.\cite{Lee} differs in important
ways from the conventional self-dual charge-density wave/superconductivity
action in one dimension. Indeed, in one dimension the displacement field
of the charge density wave is conjugate to the phase of the
superconducting order. As a result the charge density wave and the
superconducting orders are mutually exclusive (i.e. whenever
superconducting susceptibility strongly diverges the charge density wave
susceptibility does not and vice versa). In contrast, in the theory of
Ref.\cite{Lee} there exists a generic region in phase diagram where both
orders exist.

In this paper we examine in detail a one-dimensional analog of the model
studied in Ref.\cite{Lee}. The motivation for this is that well-developed
calculational methods (such as the renormalization group) can be used to
analyze the phase structure of the model. This can be used to check the
correctness of the asserted phase structure in Ref.\cite{Lee}

Now we describe the theory proposed by Lee \cite{Lee}.  Since the stripe
order is a one dimensional charge density wave, its Goldstone mode (i.e.
stripe displacement) is a $U(1)$ scalar\cite{Lee}. The superconducting
order, of course, also possesses a $U(1)$ Goldstone mode. The important
question is: how do these two $U(1)$ modes couple together? A hint of how
this coupling works comes from the experimental fact that the period of
incommensurate spin correlation decreases as the doping density increases.
Motivated by this Lee\cite{Lee} constructed the following Lagrangian
density
\begin{eqnarray}
\mathcal {L} &=& \frac{1}{2 K_{\phi\mu}} J_{\mu}^2 + \frac{1}{2
K_{\rho\mu}} q_{\mu}^2 + J_{\mu}\bar{\phi_0}\partial_{\mu}\phi_0 +
q_x(\bar{\rho_0}\partial_x\rho_0 - i g_1 J_t) \nonumber\\
&& + q_t(\bar{\rho_0}\partial_t\rho_0 - i g_2J_x). \label{lagrangian}
\end{eqnarray}
In the above $\phi_0= e^{i\theta_s}$ is the U(1) phase factor of the
superconducting order parameter, and $\rho_0=e^{i\theta_p}$ is the phase
factor of the stripe order.  That is, $\theta_p=({2\pi}/{\lambda}){\bf
\hat x}\cdot{\mathbf{u}}\mathnormal{(x,t)}$, with $\lambda$ the stripe
period and ${\mathbf{u}}(x,t)$ the displacement field of the stripe order.
$J_{\mu}$ and $q_{\mu}$ are auxiliary fields coupling to the
superconducting and stripe phases, respectively.  These auxiliary fields
have the physical interpretation of energy-momentum currents.  (In this
paper greek indices run over ${x,t}$ and repeated indices are summed.)

Without the coupling ($g_{1,2}= 0$), integrating out $J_{\mu}$ and
$q_{\mu}$ produces the field theory for two independent U(1) Goldstone
modes and their respective vortices. The effect of the coupling is to
favor stripe displacement (${\bf u}$) in the presence of local charge
imbalance ($J_t$).

To analyze \Eq{lagrangian}, Lee used a duality transformation plus an
educated guess about the four possible quantum phases corresponding to
combinations of stripe and superconducting order/disorder.  In this paper
we study a one-dimensional version of \Eq{lagrangian}, applying the
well-developed techniques of duality transformation and the
renormalization group to determine the possible phases in a more unbiased
fashion. We find that all four combinations of stripe order/disorder and
superconducting order/disorder are stable phases.  This supports the
conjectured phase structure in Ref. \cite{Lee}.

In the following we use a real-space renormalization procedure similar to
that used by Kosterlitz and Thouless to treat the phase transition of the
2-dimensional coulomb gas\cite{KT,Jose}. In section two we derive the
vortex action (the vortex of the stripe order parameter is the
dislocation). In section three we obtain the renormalization group
recursion relations for the coupling constants in that theory. As in the
Kosterlitz-Thouless theory we make the small vortex fugacity
approximation.  We analyze the implications of these flows for phase
stability in section four.

\section{Duality Transformation to 2-Species Coulomb Gas}

Following the work of Jose et al\cite{Jose} we first perform a duality
transformation and write the theory in terms of vortex degrees of freedom.

Starting with \Eq{lagrangian}, we first separate the phase of $\phi_0$ and
$\rho_0$ into a topologically trivial part and a topologically non-trivial
part:
\begin{eqnarray}
&&\phi_0=e^{i\eta_0}\phi \nonumber\\
&&\rho_0=e^{i\xi_0}\rho
\end{eqnarray}
In the above $\eta_0$ and $\xi_0$ are single valued, while $\phi$ and
$\rho$ contain configurations with non-zero windings.  After integrating
over the topologically trivial phases ($\eta_0,\xi_0$) we obtain two
conservation laws:
\begin{eqnarray}
&&\partial_{\mu}J_{\mu}=0 \nonumber \\
&&\partial_{\mu}q_{\mu}=0.
\end{eqnarray}
To explicitly fulfill these conservation laws we write
$J_{\mu}=\epsilon_{\mu\nu}\partial_{\nu}\Lambda$ and
$q_{\mu}=\epsilon_{\mu\nu}\partial_{\nu}\chi$, where $\chi$ and $\Lambda$
are scalar fields.  Substitution leads to
\begin{eqnarray}
\mathcal L & = & \frac{1}{2 K_{\rho\bar{\mu}}}(\partial_{\mu}\chi)^2 + \frac{1}{2 K_{\phi\bar{\mu}}}(\partial_{\mu}\Lambda)^2 + \epsilon_{\mu\nu}\partial_{\nu}\Lambda\bar{\phi}\partial_{\mu}\phi \nonumber\\
&&+ \epsilon_{\mu\nu}\partial_{\nu}\chi\bar{\rho}\partial_{\mu}\rho +
ig_1\partial_t\chi\partial_x\Lambda + ig_2\partial_x\chi\partial_t\Lambda.
\end{eqnarray}
Upon integrating by parts and identifying the vortex densities
$N=i\epsilon_{\mu\nu}\partial_{\nu}(\bar{\rho}\partial_{\mu}\rho)$ and
$M=i\epsilon_{\mu\nu}\partial_{\nu}(\bar{\phi}\partial_{\mu}\phi)$ the
Lagrangian density becomes

\begin{eqnarray}
\mathcal L &=& \frac{1}{2 K_{\rho\bar{\mu}}}(\partial_{\mu}\chi)^2 + \frac{1}{2 K_{\phi\bar{\mu}}}(\partial_{\mu}\Lambda)^2 +i \Lambda M +i \chi N \nonumber\\
&& - i(g_1+g_2)\Lambda\partial_t\partial_x\chi.
\end{eqnarray}

\noindent The above equation can be written in momentum space as

\begin{eqnarray}
\cal L &=& \frac{1}{2}\left(\!\! \begin{array}{cc}\chi({\bf k})&\!\!\!\Lambda({\bf k})\end{array}\!\!\right)^{\ast} \left(\!\! \begin{array}{cc}\frac{k_{\mu}^2}{K_{\rho\bar{\mu}}} &iGk_xk_t\\iGk_xk_t &\frac{k_{\mu}^2}{K_{\phi\bar{\mu}}}\end{array}\!\!\right)\left(\!\!\!\begin{array}{c}\chi({\bf k})\\\Lambda({\bf k})\end{array}\!\!\!\right)\nonumber\\
&&\qquad{} + i \left(\!\! \begin{array}{cc}\chi({\bf
k})&\!\!\!\Lambda({\bf k})\end{array}\!\!\right)^{\ast}\left(\!\!\!
\begin{array}{c}N({\bf k})\\M({\bf k})\end{array}\!\!\!\right)
\end{eqnarray}
where $G=g_1+g_2$. Integrating out the $\chi$ and $\Lambda$ fields then
produces

\begin{eqnarray}
\cal L &=&  \frac{1}{2} \left(\!\! \begin{array}{cc}N&\!\!\!M\end{array}\!\! \right)^{\ast} \frac{1}{det} \left(\!\! \begin{array}{cc} \frac{k_{\mu}^2}{K_{\phi\bar{\mu}}} &-iGk_xk_t\\-iGk_xk_t &\frac{k_{\mu}^2}{K_{\rho\bar{\mu}}}\end{array}\!\!\right)\!\left(\!\!\!\begin{array}{c}N\\M\end{array}\!\!\!\right) \label{ours}\\
det&=&\left(\frac{k_{\mu}^2}{K_{\rho\bar{\mu}}}\right)\left(\frac{k_{\nu}^2}{K_{\phi\bar{\nu}}}\right)
+ G^2k_x^2k_t^2 .\label{det}
\end{eqnarray}
\Eq{det} is the starting point of our renormalization group analysis. It
describes a system of two interacting (anisotropic) coulomb gases -- the
vortices of the superconducting order parameter and the dislocations of
the stripe order parameter.  Inspired by the work of Kosterlitz and
Thouless we perform a real space renormalization group analysis of
\Eq{det} in the following.

\section{Renormalization Group Analysis}

\Eq{det} is more complicated than the one species Coulomb gas problem in
two respects: 1) there are two species of vortices and 2) the interactions
are not rotationally invariant (i.e. the interaction depends not only on
the distance between vortices but also on their relative orientation). In
order to complete the renormalization group program we have to
characterize the interaction in terms of a discrete set of coupling
constants. One way of achieving this is to Fourier transform the angular
dependence of the vortex-vortex interaction.  In momentum space each
element of the interaction matrix is of the form $G({\bf k}
)=G(k,\theta)=\frac{g(\theta)}{k^2}$ ($\theta$ is the angle made by ${\bf
k}$ and the $k_x$ axis).  Therefore we expand each of these terms in a
fourier series, {\it e.g.} $g(\theta) = \sum_n a_n e^{(in\theta)}$. When
transformed back to real space, our action then becomes

\begin{eqnarray}
\cal S &=& \frac{1}{2}\int\!d^2{\bf R}_1 \,d^2{\bf R}_2 \,N({\bf R}_1 ) \:G_N({\bf R}_1 - {\bf R}_2 ) \: N({\bf R}_2 )
\nonumber\\
&&+ M({\bf R}_1 ) \: G_M({\bf R}_1 - {\bf R}_2 ) \: M({\bf R}_2 )\nonumber\\
&&+ 2 M({\bf R}_1 ) \: i\Gamma( {\bf R}_1 -{\bf R}_2 ) \: N({\bf
R}_2 ), \label{action}
\end{eqnarray}
where
\begin{eqnarray}
G_N ({\bf R})&=& \int\!\frac{d^2{\bf k} }{(2\pi)^2} \left(\sum_{n=-\infty}^{\infty} a_n e^{in\theta}\right) \frac{e^{i {\bf k}\cdot{\bf R}}}{k^2}\label{GN}\\
G_M ({\bf R})&=& \int\!\frac{d^2{\bf k} }{(2\pi)^2} \left(\sum_{n=-\infty}^{\infty} \alpha_n e^{in\theta}\right) \frac{e^{i{\bf k}\cdot{\bf R}}}{k^2}\label{GM}\\
\Gamma ({\bf R})&=& \int\!\frac{d^2{\bf k} }{(2\pi)^2}
\left(\sum_{n=-\infty}^{\infty} c_n e^{in\theta}\right) \frac{e^{i{\bf
k}\cdot {\bf R}}}{k^2}\label{Gamma}
\end{eqnarray}
In the above
\begin{eqnarray}
&&a_n=(-1)^n a_{-n}^{\ast}\nonumber\\
&&\alpha_n=(-1)^n \alpha_{-n}^{\ast}\nonumber\\
&&c_n=(-1)^n c_{-n}^{\ast}\label{cons1}
\end{eqnarray}
to ensure the interaction functions are real. We stress that because of
the angular dependence of Eqs. \ref{GN}, \ref{GM}, and \ref{Gamma}, $G_N,
G_M$ and $\Gamma$ depend not only on the distance $|{\bf R_1}-{\bf R_2}|$
but also on the relative orientation  $({\bf R_1}-{\bf R_2})/|{\bf
R_1}-{\bf R_2}|$. In general $a_n$ and $\alpha_n$ are nonzero only for
even $n$.  The physical reason for this is indistinguishability of two
charges of the same type (for details see the appendix).  $c_n$ can be
nonzero for both odd and even $n$.

\begin{figure}
\includegraphics[width=8.5cm,height=6cm]{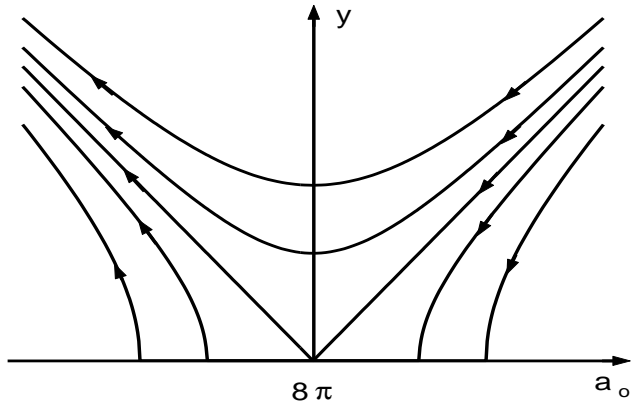}
\caption{Renormalization group flow for the Kosterlitz-Thouless
transition. The fixed points are the y=0 axis.  In our model this
corresponds to only $y$, $a_0$ nonzero.} \label{KTflow}
\end{figure}

The limit where all the $a_n$, $\alpha_n$, and $c_n$ are zero except $a_0$
and $\alpha_0$ describes two decoupled isotropic 2-D X-Y models in their
coulomb gas representations -- the Kosterlitz and Thouless problem. Before
we attack \Eq{det}, as a warm up, let us briefly review the
Kosterlitz-Thouless results for the one component system. In the
renormalization group approach one integrates out one pair of tightly
bound dipole (i.e. a dipole with $r_c+dr_c < size < r_c$) at a time. The
renormalization group proceeds iteratively by treating $r_c$ as a running
length scale. The two coupling constants in that case are  the
vortex-vortex interaction strength $a_0$ and vortex fugacity $y= e^{-\mu}$
where $\mu$ is the core energy of vortices.  In the limit of $y<<1$, the
renormalization group equations for $a_0$ and $y$ are given by
\begin{eqnarray}
\frac{dy}{dl} &=& y\left( 2-\frac{a_0}{4\pi} \right)\\
\frac{da_0}{dl} &=& -\pi y^2a_0^2.\label{ord}
\end{eqnarray}
The above equations have the entire $y=0$ axis as fixed points. However
depending on whether $a_0-8\pi$ is positive/negative the fixed point is
stable/unstable. The point $y=0, a_0=8\pi$ is a critical point.  Near it,
the flow trajectories are given by
\begin{equation}
a_0^2 - (4 \pi)^4 y^2 = C \label{trajectory}
\end{equation}
Here C is a constant labelling each trajectory.  This flow is shown in
figure \ref{KTflow}. Note that the $C=0$ separatrix $y=\frac{1}{(4\pi)^4}
(a_0-8\pi)$ separates the basins of attraction for the ordered and
disordered phases. In the ordered phase the density of vortices
renormalizes to zero ($y \rightarrow 0$) at large length scales,
signifying the presence of a bound dipole phase. In the disordered phase
the density of vortices increases ($y$ increases) at large length scales,
signifying the existence of a vortex plasma (unbound dipole) phase.

The two-component vortex gas problem we are facing is not so different.
However, we have to painstakingly keep track of all the Fourier
coefficients in \Eq{GN}, \Eq{GM} and \Eq{Gamma} and examine how they
renormalize. Interestingly, even in the presence of these anisotropic
interactions the Kosterlitz-Thouless renormalization group program closes.

All the technical details are given in the appendix. Here we just note the
following point. Because the ``Coulomb charge'' of the superconducting
vortices is not related to the ``Coulomb charge'' of the stripe vortices,
only intra-species dipoles are possible.  This implies that the positions
of vortices belonging to different species do not have to obey the
constraint that the minimum distance is $r_c$. To lowest order in $y$ the
resulting renormalization group equations are given by
\begin{eqnarray}
\frac{dy_N}{dl} &=& y_N\left( 2 - \frac{a_0}{4\pi} \right)\nonumber\\
\frac{dy_M}{dl} &=& y_M\left( 2 - \frac{\alpha_0}{4\pi} \right)\nonumber\\
\frac{da_n}{dl} &=& -\pi \sum_{k,m} \delta_{n,k+m} \left( y_N^2 a_k a_m + y_M^2 (-1)^{m+1} c_k c_m\right)\nonumber\\
\frac{d\alpha_n}{dl} &=& -\pi \sum_{k,m} \delta_{n,k+m} \left(y_M^2 \alpha_k \alpha_m + y_N^2 (-1)^{m+1} c_k c_m\right)\nonumber\\
\frac{dc_n}{dl} &=& -\pi \sum_{k,m} \delta_{n,k+m} \left(y_N^2 a_k
+ y_M^2 \alpha_k \right) c_m.  \label{flowend}
\end{eqnarray}

Note that the renormalization of $y_{N,M}$  only depends on the isotropic
part of the coupling  ($a_0$ and $\alpha_0$). The renormalization of an
anisotropy coefficient, ({\it e.g.} $a_n$) includes many terms. Each term
is a quadratic function of $a_k$, $\alpha_k$, or $c_k$. If we set all
coupling constants except $a_0$ and  $\alpha_0$ to zero we recover the
Kosterlitz-Thouless flow equations (for two separate species). It is
easily verified that the condition for $G_N, G_M$, and $\Gamma$ to be real
($a_n=(-1)^n a_{-n}^{\ast}$, etc.) is preserved by these flow equations.
It is also clear from the form of these equations that these coefficients
form a closed set under renormalization.

\section{Phases of the Two-Species Coulomb Gas}

\Eq{flowend} predicts fixed points for $y_M = y_N = 0$ and $a_n$,
$\alpha_n$, $c_n =$ anything.  As in the normal Kosterlitz-Thouless case,
we interpret $y=0$ as the absence of unbound dipoles. From first line of
\Eq{flowend}, $y_N=0$ is linearly stable if $a_0>8\pi$. For $a_0>8\pi$ the
renormalization group brings $y_N$ to larger values. Similar statements
hold for $\alpha_0$ and $y_M$.  This suggests the presence of four phases
depending on whether the vortices of $N$ or $M$ species form dipoles or
unbind.

However, this is not quite enough for our purposes. What we really need to
know is whether all four phases can be reached by varying the five
parameters in \Eq{ours}.  Put another way, the physical system of
\Eq{ours} lives in a five dimensional subspace of the infinite dimensional
space formed by the $a_n$, $\alpha_n$, and $c_n$.  We need to check which
phases can be reached by trajectories originating in the physical
subspace, not just which phases exist for the infinite dimensional space.

In order to obtain a tractable problem, in the following we concentrate on
the case in which $K_{\rho\mu} = K_{\rho}$, $K_{\phi\mu} = K_{\phi}$ and
$GK_{\rho,\phi}<<1$.  When $GK_{\rho,\phi}$ are small it is easy to
evaluate the fourier coefficients in equations Eqs. \ref{GN}, \ref{GM},
and \ref{Gamma} in powers of $GK_{\rho,\phi}$. If $GK_{\rho,\phi} = {\cal
O}(\epsilon)$ we find that the leading contribution to $a_n$, $\alpha_n$,
$c_n$ is ${\cal O}(\epsilon^{\vert n \vert/2})$. For the specific form of
interaction in \Eq{ours} it is simple to see that besides Equations
\ref{cons1} there are additional constraints on $a_n$,$\alpha_n$ and $c_n$
:
\begin{eqnarray}
&&a_n=\alpha_n=0~~{\rm unless}~ n=4m \nonumber \\
&& c_n=0~~{\rm unless}~n= 4m + 2.\label{cons2}
\end{eqnarray}
All of these conditions are preserved by the flow equations.  In terms of
the original parameters in \Eq{ours}, the non-vanishing coefficients up to
order $\epsilon$ are:
\begin{eqnarray}
&&a_0 = K_{\rho}(1-K_{\phi} K_{\rho} G^2/8) \nonumber\\
&&\alpha_0 = K_{\phi}(1-K_{\phi} K_{\rho} G^2/8) \nonumber\\
&&c_2 = c_{-2}^{\ast} = i G K_{\rho} K_{\phi}/4 \label{coeff}
\end{eqnarray}
In the following we truncate the space considered to only these
coefficients, which is correct to lowest order in $\epsilon$. Furthermore,
this restricts us to a five-dimensional space of parameters, which we can
take to be independently determined by the five parameters in \Eq{ours}.

In this case, the flow equations \ref{flowend} become
\begin{eqnarray}
\frac{dy_N}{dl} &=& y_N\left( 2 - \frac{a_0}{4\pi} \right)\label{yneqn} \\
\frac{dy_M}{dl} &=& y_M\left( 2 - \frac{\alpha_0}{4\pi} \right) \label{ymeqn}\\
\frac{da_0}{dl} &=& -\pi \left(y_N^2 a_0^2 - y_M^2 \vert c_2 \vert^2 \right) \label{aeqn} \\
\frac{d\alpha_0}{dl} &=& -\pi \left(y_M^2 \alpha_0^2 - y_N^2 \vert c_2 \vert^2 \right) \label{alphaeqn} \\
\frac{dc_2}{dl} &=& - \pi \left(y_N^2 a_0 c_2 + y_M^2 \alpha_0 c_2
\right)\label{ceqn}
\end{eqnarray}

First, by multiplying \Eq{ceqn} by $c_2^{\ast}$ and then adding the result
to its complex conjugate, we obtain
\begin{equation}
\frac{d\vert c_2\vert^2}{dl} = - 2\pi \left(y_N^2 a_0 + y_M^2 \alpha_0
\right) \vert c_2 \vert^2 \label{csqr}
\end{equation}
Using this in \Eq{aeqn} and \Eq{alphaeqn} then gives us
\begin{eqnarray}
\frac{da_0}{dl} &=& -\pi y_N^2 \left(a_0^2 + \vert c_2 \vert^2 \right) - \frac{1}{16 \pi} \frac{d \vert c_2 \vert^2}{dl} \\
\frac{d\alpha_0}{dl} &=& -\pi y_M^2\left(\alpha_0^2 - \vert c_2 \vert^2 \right) - \frac{1}{16 \pi} \frac{d \vert c_2 \vert^2}{dl}\\
\end{eqnarray}

At this point, we examine closely the region of parameter space around the
critical point by making the change of variables
\begin{eqnarray*}
a &=& a_0 - 8\pi \\
\alpha &=& \alpha_0 - 8\pi \\
c &=& c_2 - {\bar c}
\end{eqnarray*}
In the above ${\bar c}$ is the fixed point of $c_2$.  After some algebra,
and keeping terms to lowest order in $a$, $\alpha$, $c$, $y_N$, and $y_M$,
the flow equations for $a$ and $\alpha$ are, after some algebra,
\begin{eqnarray}
\frac{dy_N^2}{dl} &=& \frac{1}{2\pi} y_N^2 a \\
\frac{dy_M^2}{dl} &=& \frac{1}{2\pi} y_M^2 \alpha \\
\frac{da^2}{dl} &=& -2\pi y_N^2 a\left((8\pi)^2 + \vert {\bar c} \vert^2 \right) - \frac{a}{16 \pi} \frac{d \vert c_2 \vert^2}{dl} \label{mid3}\\
\frac{d\alpha^2}{dl} &=& -2\pi y_M^2 \alpha \left((8\pi)^2 - \vert {\bar c} \vert^2 \right) - \frac{\alpha}{16 \pi} \frac{d \vert c_2 \vert^2}{dl}\label{mid4}\\
\end{eqnarray}

\begin{figure}
\includegraphics[width=8.5cm, height=6cm]{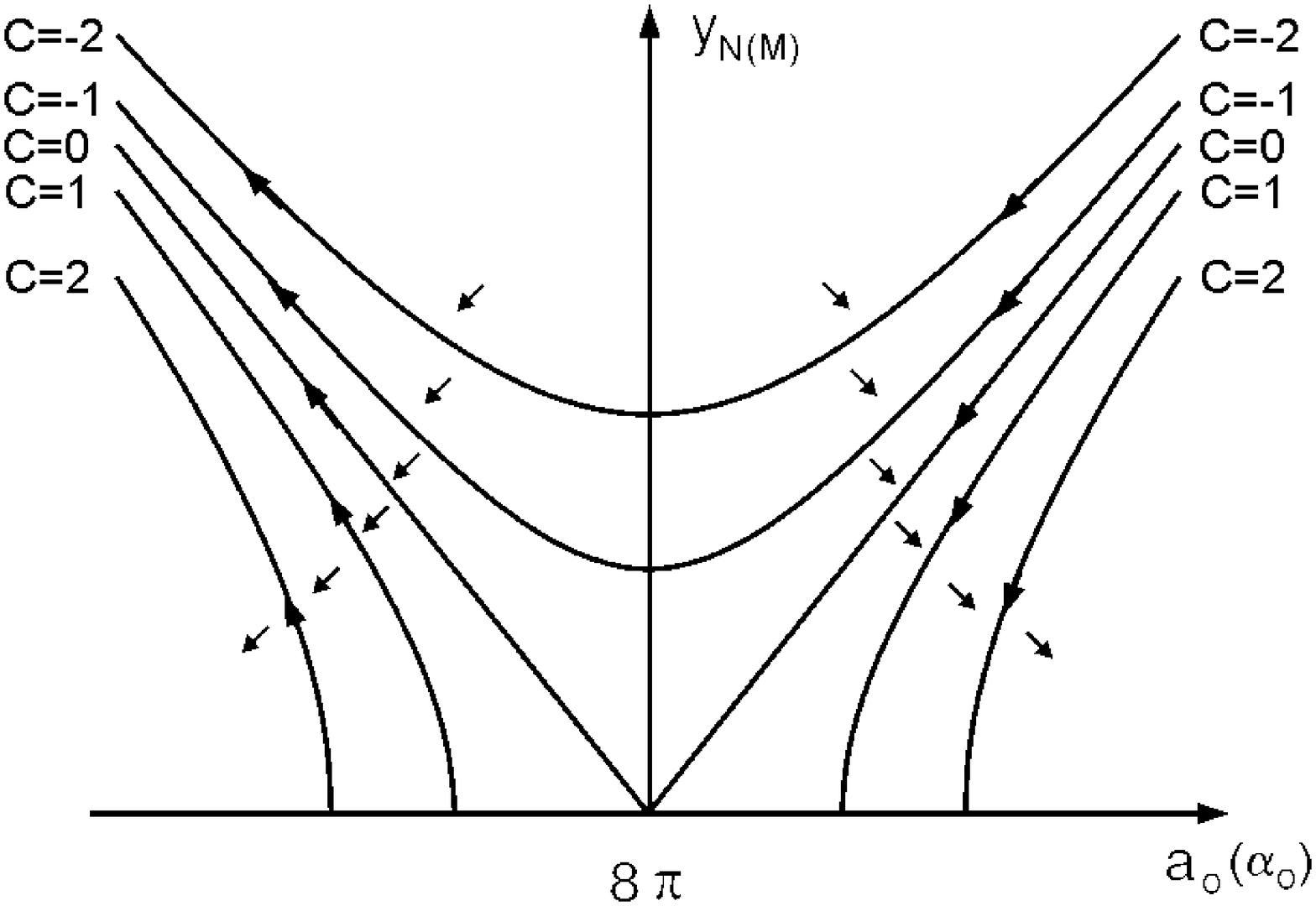}
\caption{Renormalization group flow for weakly coupled
superconducting(stripe) vortex gases, with contour numbers $C$ labelled.
The fixed points are the y=0 axis. With nonzero coupling, the flow moves
from one contour to the next as indicated by the arrows.  It is clear that
even in the presence of coupling for each gas there are still two phases
corresponding to $y \rightarrow 0$ and $y$ increasing.}
\label{interactingflow}
\end{figure}

Finally, we can combine these equations to yield
\begin{eqnarray}
&&\frac{d \left( a^2 - X y_N^2 \right)}{dl} = \frac{d C_N}{d l} = -
\frac{a}{8\pi}\frac{d \vert
c_2 \vert^2}{dl} \label{Ncontour}\\
&&\frac{d \left( \alpha^2 - X y_M^2 \right)}{dl} = \frac{d C_M}{d l} = -
\frac{\alpha}{8\pi}\frac{d \vert c_2 \vert^2}{dl} \label{Mcontour}
\end{eqnarray}
Here $X= (4 \pi)^4 (1 + \vert {\bar c} \vert^2/(8 \pi)^2)$.  To understand
these equations, note that the quantity in parentheses on the LHS of each
equation is precisely of the form of the contour numbers for trajectories
in the Kosterlitz-Thouless case (cf. \Eq{trajectory}), with the slope of
the separatrix renormalized from $\frac{1}{(4\pi)^4}$ to $1/X$. In the
absence of interactions ($c=0$) these contour numbers $C_N$ and $C_M$ are
conserved, but in the presence of interactions the renormalization group
pushes the flow from one contour to the next.  Furthermore, by
\Eq{Ncontour} and \Eq{csqr}, for $a>0$ (i.e. $a_0 > 8\pi$) the contour
number $C_N$ increases, while for $a<0$ (i.e. $a_0 < 8\pi$) $C_N$
decreases. Similarly for $C_M$ and $\alpha$.  The resulting flow is
diagrammed in figure \ref{interactingflow}.

It is clear from this that for a trajectory originating below the
separatrices ($y_N < (a_0 - 8\pi)/X$ and $y_M < (\alpha_0 - 8\pi)/X$), the
flow leads to both $y_N$ and $y_M$ zero.  In other words, there is a
stable phase with both types of vortices bound as dipoles, corresponding
to a stripe ordered/superconducting ordered phase.  If the trajectory
starts with $(a_0 - 8\pi) < 0$ and $(\alpha_0 - 8\pi) < 0$, the flow leads
to both $y_N$ and $y_M$ increasing and unbound dipoles of both species.
Thus there is a stripe disordered/superconducting disordered phase.
Finally, in the mixed case, {\it e.g.} $y_N < (a_0 - 8\pi)/X$ and
$(\alpha_0 - 8\pi) < 0$, $y_N$ flows to zero but $y_M$ increases.  Thus
there are phases with stripe disorder/superconducting order and stripe
order/superconducting disorder.

\begin{figure}
\includegraphics[width=8.5cm, height=6cm]{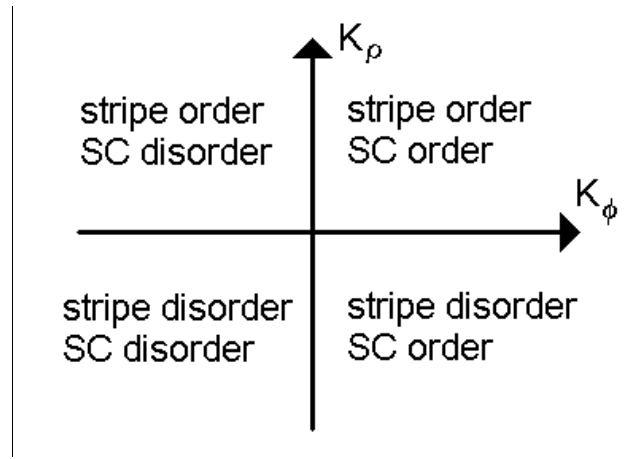}
\caption{Phase diagram for interacting stripe and Superconducting order,
for isotropic couplings $K_{\phi}$ and $K_{\rho}$ and weak interaction
$G$. Only the $K_{\phi}-K_{\rho}$ plane is shown.  Although different
values of $G$ correspond to different long-range interactions, $G$ does
not control the existence of stripe/superconducting order or disorder.}
\label{phasediagram}
\end{figure}

We have seen that, with weak $(g_1+g_2)$ all four phases corresponding to
stripe order/superconducting order, stripe disorder/superconducting order,
stripe order/superconducting disorder, and stripe disorder/superconducting
disorder are stable and can be realized in the system described by Lee's
theory in (1+1) dimensions (\Eq{lagrangian}).  We emphasize that we have
analyzed only the case with isotropic couplings $K_{\rho\mu} = K_{\rho}$
and $K_{\phi\mu} = K_{\phi}$ and weak coupling. In this case, the phase
diagram in the $K_{\phi}/K_{\rho}$ plane is shown in Fig.
\ref{phasediagram}.

\section{Summary}
The main results of this paper are the interacting coulomb gas
representation of the competing stripe and superconducting orders,
\Eq{det}; and the renormalization group flow \Eq{flowend}. Analysis of
these flow equations shows that the 1+1 dimensional version of the theory
proposed in ref. \cite{Lee} supports stable phases corresponding to stripe
order/superconducting order, stripe disorder/superconducting order, stripe
order/superconducting disorder, and stripe disorder/superconducting
disorder.

{\bf Acknowledgments:} I thank Qiang-Hua Wang and Dung-Hai Lee for useful
discussions and comments on the manuscript.  H.F. was supported in part by
the Department of Defense through a National Defense Science and
Engineering Graduate Fellowship.

\appendix
\section{Details of renormalization group calculation}

Here we present the details for the renormalization of the system
described by \Eq{ours} and \Eq{det}, and parametrized in fourier
coefficients via equations \ref{GN}, \ref{GM}, and \ref{Gamma}.  The
details closely follow the procedure used by Jose {\it et.
al.}\cite{Jose}.  The basic idea is to introduce a small length scale
cutoff $r_c$, and then integrate out configurations with pairs of the same
type of charge which are $r_c + dr_c$ apart to find a new system with a
longer minimum length scale.

In our action \Eq{action}, the fields $N$ and $M$ consist of point
charges, {\it e.g.} $N({\bf r}) = \sum_{\alpha} N_{\alpha} \delta({\bf r}-
{\bf r}_{\alpha})$.  We will express the terms \Eq{action} involving $G_N$
and $G_M$ as sums over pairs of these charges.  This results in the
cancellation of all odd fourier components of $G_N$ and $G_M$, {\it i.e.}
$a_n$ and $\alpha_n$ are zero for odd $n$. To see this, note that in
re-expressing the sum as a sum over pairs, we combine terms like
$N_{\alpha}N_{\beta} G_N({\bf R}_{\alpha} - {\bf R}_{\beta} ) +
N_{\beta}N_{\alpha} G_N({\bf R}_{\beta} - {\bf R}_{\alpha} )$.  The two
$G_N$'s above differ by reversing the relative vector (i.e.
$\theta\rightarrow\theta+\pi$).  Since odd fourier components pick up a
relative minus sign under this reversal, they cancel each other, and only
the even fourier components survive.  On the other hand, since the terms
with $\Gamma$ describe the interaction between distinguishable vortices,
we cannot convert them into sums over pairs, and therefore the fourier
coefficients $c_n$ can be nonzero for both odd and even $n$.

As in the Kosterlitz-Thouless case, $G_N$ and $G_M$ diverge
logarithmically at short length scales.  In \Eq{ours} this translates to
divergences when ${\bf R}_1 = {\bf R}_2$.  To remove this divergence, we
must enforce charge neutrality ($\sum_{\alpha} N_{\alpha} = \sum_{\alpha}
M_{\alpha} = 0$) and impose a small distance cut-off $r_c$\cite{Kogut}. To
account for the microscopic physics lost in this procedure, we introduce
core energies $\Delta_N$ and $\Delta_M$ for the vortices (charges).  After
this our action is written as

\begin{widetext}

\begin{eqnarray}
\cal S &=& \sum_{(\alpha,\beta)} N_{\alpha} N_{\beta} G_N({\bf
R}_{\alpha}-{\bf R}_{\beta}) + \sum_{(\alpha,\beta)} M_{\alpha}
M_{\beta} G_M({\bf R}_{\alpha}-{\bf R}_{\beta}) + \sum_{\alpha,
\beta} M_{\alpha} N_{\beta} i\Gamma_N({\bf R}_{\alpha}-{\bf
R}_{\beta}) + \sum_{\alpha} N_{\alpha}^2 \Delta_N +\sum_{\alpha}
M_{\alpha}^2 \Delta_M \nonumber\\
\end{eqnarray}
where $(\alpha,\beta)$ denotes a sum over pairs and $\alpha,\beta$ denotes
an unrestricted sum over both $\alpha$ and $\beta$.  At this point we make
the simplifying assumption that $\Delta_N$ and $\Delta_M$ are very large,
so we may restrict to $N_{\alpha}, M_{\alpha} = \pm 1$.  Introducing the
fugacities $y_N = e^{-\Delta_N}$ and $y_M = e^{-\Delta_M}$, we can write
the partition function as a sum over $j$ N-dipoles and $k$ M-dipoles:
\begin{eqnarray}
Z &=& \sum_{j=0}^{\infty} \sum_{k=0}^{\infty} y_N^{2j} y_M^{2k} \frac{1}{j!^2}\frac{1}{k!^2} \int\!\frac{d^2{\bf x}_1}{r_c^2}\cdots\frac{d^2{\bf x}_{2j}}{r_c^2} \int\!\frac{d^2{\bf z}_1}{r_c^2}\cdots\frac{d^2{\bf z}_{2k}}{r_c^2} e^{-\tilde {S}}\label{Z}\\
\tilde{S}&=& \sum_{(\alpha,\beta)} N_{\alpha} N_{\beta} G_N({\bf
x_{\alpha}}-{\bf x}_{\beta}) + \sum_{(\alpha,\beta)} M_{\alpha} M_{\beta}
G_M({\bf z}_{\alpha}-{\bf z}_{\beta}) + \sum_{\alpha, \beta} M_{\alpha}
N_{\beta} i\Gamma_N({\bf z}_{\alpha}-{\bf x}_{\beta}) \label{S}
\end{eqnarray}

The first step in the real-space renormalization procedure is to integrate
over dipoles of size $r_c + dr_c$.  Because $y_N, y_M \ll 1$ we need only
consider configurations with one dipole of either type.  Furthermore,
mixed M-N dipoles are not integrated out because doing so would violate
overall charge neutrality for each species.  This means that the M and N
charges are configured freely with respect to each other.  Writing out
these single dipole contributions, the partition function gains an extra
factor:
\begin{eqnarray}
e^{-\tilde {s}} & \rightarrow & e^{-\tilde {s}} \times \left\{1 + \int\!\frac{d^2{\bf R}_N}{r_c^2}\int^{r_c+dr_c}_{r_c}\frac{d^2{\bf r}_N}{r_c^2}\,y_N^2 e^{-S'_N} +  \int\!\frac{d^2{\bf R}_M}{r_c^2}\int^{r_c+dr_c}_{r_c}\frac{d^2{\bf r}_M}{r_c^2}\,y_M^2 e^{-S'_M}  + {\cal O}(y^4) \right\} \label{A4}\\
S'_N &=& -G_N({\bf r}_N) + \sum_{\alpha}^{2k}N_{\alpha}\left( G_N({\bf R}_N + \frac{{\bf r}_N}{2} - {\bf x}_{\alpha}) - G_N({\bf R}_N - \frac{{\bf r}_N}{2} - {\bf x}_{\alpha}) \right) +\nonumber\\
&& \sum_{\alpha}^{2j}M_{\alpha}\left( i\Gamma({\bf R}_N + \frac{{\bf r}_N}{2} - {\bf z}_{\alpha}) - i\Gamma({\bf R}_N - \frac{{\bf r}_N}{2} - {\bf z}_{\alpha}) \right) \label{A5}\\
S'_M &=& -G_M({\bf r}_M) + \sum_{\alpha}^{2j} M_{\alpha}\left( G_M({\bf R}_M + \frac{{\bf r}_M}{2} - {\bf z}_{\alpha}) - G_M({\bf R}_M - \frac{{\bf r}_M}{2} - {\bf z}_{\alpha}) \right) +\nonumber\\
&& \sum_{\alpha}^{2k}N_{\alpha}\left( i\Gamma({\bf R}_M + \frac{{\bf
r}_M}{2} - {\bf x}_{\alpha}) - i\Gamma({\bf R}_N - \frac{{\bf r}_N}{2} -
{\bf x}_{\alpha}) \right) \label{actionprime}
\end{eqnarray}
where ${\bf R}$ signifies the center and ${\bf r}$ the separation of the
dipole being integrated out.  To obtain the above, we have used the fact
that the combinatorial factor for $j+1$ pairs of charges, one of which is
a dipole ({\it i.e.} $\frac{j+1)^2}{(j+1)!^2}$), is equal to the
combinatorial factor of j pairs of charges({\it i.e.} $\frac{1}{j!^2}$).

To proceed, rewrite equations \ref{A5} and \ref{actionprime} using
identities like $\left( G_N({\bf R}_N + \frac{{\bf r}_N}{2} - {\bf
x}_{\alpha}) - G_N({\bf R}_N - \frac{{\bf r}_N}{2} - {\bf x}_{\alpha})
\right) = {\bf r}_N\cdot\nabla_{{\bf R}_N} G_N({\bf R}_N - {\bf
x}_{\alpha})$.  In \Eq{A4}, these expressions enter into exponentials,
which we expand.  The linear terms in this expansion produce nothing
interesting.  To see this, note that there are two types of these terms:
those with a gradient, and those without.  The linear terms involving the
gradient do not contribute since their integral with respect to ${\bf r}$
is zero. The linear terms involving no gradient ($-G_N({\bf r}_N)$ and
$-G_M({\bf r}_M)$) integrate to constants independent of $j$, $k$, and the
$N_{\alpha}$, $M_{\alpha}$, so they merely produce an overall constant
multiplying the partition function.

Since the linear terms in the expansion of the exponentials of \Eq{A4} are
uninteresting, we look at the second order terms.  There are three types
of terms here: those involving no gradients (e.g. $G_N G_N$); those
involving one gradient (e.g. $G_N {\bf r} \cdot \nabla G_N$); and those
involving two gradients (e.g. ${\bf r}\cdot\nabla G_N {\bf r}\cdot\nabla
G_N$).  The first type of term, with no gradients, integrates to a
constant independent of $j$, $k$, and the $N_{\alpha}$, $M_{\alpha}$, so
we ignore it.

The second type of term, with only one gradient, integrates to zero. For
example, one such expression is $\sum_{\alpha} N_{\alpha} G({\bf r}_N)
{\bf r}_N \cdot \nabla_{{\bf R}_N} G_N({\bf R}_N - {\bf x_{\alpha}})$.
However, in each term of the sum we can change integration variables to
${\bf R}'_N =  {\bf R}_N - {\bf x_{\alpha}}$, producing $\int
\sum_{\alpha} N_{\alpha} G({\bf r}_N) {\bf r}_N \cdot \nabla_{{\bf R}'_N}
G_N({\bf R}'_N )$. Charge neutrality then makes the sum over $N_{\alpha}$
zero.

The third type of term, with two gradients, does contribute to
renormalization.  A typical term of this type is (using the fourier
expansion of $\Gamma$)
\begin{eqnarray*}
\frac{1}{2!}\int\!\frac{d^2{\bf R}_N}{r_c^2}\int^{r_c+dr_c}_{r_c}\frac{d^2{\bf r}_N}{r_c^2}\,y_N^2 \sum_{\alpha, \beta} M_{\alpha}M_{\beta} {\bf r}_N\cdot\nabla_{{\bf R}_N}\left[ i\int\!\frac{d^2{\bf k}}{(2\pi)^2}\frac{1}{k^2} \left( \sum c_n e^{in\theta}\right) e^{i {\bf k}\cdot({\bf z}_{\alpha}-{\bf R}_N)} \right]\times\\
{\bf r}_N\cdot\nabla_{{\bf R}_N}\left[ i\int\!\frac{d^2{\bf
k}}{(2\pi)^2}\frac{1}{k^2} \left( \sum c_m e^{im\theta}\right) e^{i {\bf
k}\cdot({\bf z}_{\beta}-{\bf R}_N)} \right]
\end{eqnarray*}
Using ${\bf r}_N \approx r_c(\cos \phi, \sin \phi)$ we can integrate over
${\bf r}_N$ to obtain
\begin{displaymath}
-\frac{1}{2!}\frac{2\pi}{2}\frac{r_c^3dr_c}{r_c^2}y_N^2\int\!\frac{d^2{\bf
R}_N}{r_c^2}\sum_{\alpha, \beta}M_{\alpha}M_{\beta} \int\!\frac{d^2{\bf
k'}}{(2\pi)^2}\frac{d^2{\bf k}}{(2\pi)^2} (-i{\bf k})\cdot(-i{\bf
k'})\frac{1}{k^2k^{\prime 2}} \sum_{n,m} c_n c_m
e^{in\theta+im\theta'}e^{i{\bf k}\cdot{\bf z}_{\alpha} + i{\bf
k'}\cdot{\bf z}_{\beta}}e^{-i({\bf k} +{\bf k'})\cdot{\bf R}_N}
\end{displaymath}
\begin{displaymath}
= -\frac{\pi}{2!}\frac{dr_c}{r_c}y_N^2
\sum_{\alpha,\beta}M_{\alpha}M_{\beta} \int\!\frac{d^2{\bf k}}{(2\pi)^2}
\frac{1}{k^2}e^{i{\bf k}\cdot({\bf z}_{\alpha} - {\bf z}_{\beta})}
\sum_{n,m} c_n c_m e^{i(n+m)\theta}e^{im\pi}
\end{displaymath}
\begin{displaymath}
= \frac{\pi}{2}\frac{dr_c}{r_c}y_N^2
\sum_{\alpha,\beta}M_{\alpha}M_{\beta} \int\!\frac{d^2{\bf k}}{(2\pi)^2}
\frac{1}{k^2}e^{i{\bf k}\cdot({\bf z}_{\alpha} - {\bf z}_{\beta})} \sum_n
\left(\sum_{l,m} \delta_{n,l+m} c_l c_m (-1)^{m+1}\right) e^{in\theta}
\end{displaymath}
\end{widetext}

After many calculations like this and some algebra, we obtain a new
partition function in the form of \Eq{Z} again, except that the
interaction functions have been redefined to $G'_N$, $G'_M$, and $\Gamma'$
via the new coefficients
\begin{eqnarray}
a'_n &=& a_n - \pi \sum_{k,m} \delta_{n,k+m}(y_N^2 a_k a_m + y_M^2(-1)^{m+1} c_k c_m) dl\nonumber\\
\alpha'_n &=& \alpha_n - \pi \sum_{k,m}\delta_{n,k+m}(y_M^2 \alpha_k \alpha_m + y_N^2(-1)^{m+1} c_k c_m) dl\nonumber\\
c'_n  &=& c_n - \pi \sum_{k,m} \delta_{n,k+m}(y_N^2 a_k + y_M^2 \alpha_k)
c_m dl. \label{c}
\end{eqnarray}
Here $dl = dr_c/r_c$.

At this point it is useful to examine more closely the structure of the
contributions to the interaction functions and understand what sorts of
interactions we are dealing with. To this end we evaluate a typical term,

\begin{equation}
\int\!\frac{d^2{\bf k} }{(2\pi)^2} e^{in\theta}\frac{e^{i{\bf k} \cdot{\bf
r} }}{k^2}\nonumber
\end{equation}
using ${\bf k} \cdot {\bf r}=kr\cos\phi$ this becomes
\begin{equation}
\frac{1}{(2
\pi)^2}\int^{2\pi}_0\!d\theta\int^{\infty}_\frac{1}{L}\frac{dk}{k}
e^{ikr\cos(\theta-\phi)}e^{in\theta}
\end{equation}
\begin{equation}
 = \frac{i^n e^{in\phi}}{2\pi}\int_\frac{r}{l}^\infty \frac{dx}{x} J_{-n}(x)
\end{equation}
where we have introduced the Bessel function $J_n$.  Due to the
oscillatory nature of the Bessel function the integral is dominated by the
infrared so we employ the asymptotic form $J_n(x) \sim
\frac{1}{n!}(\frac{x}{2})^n$ to obtain
\begin{equation}
\frac{(\pm i)^n}{2\pi n!} e^{in\phi}\left\{ \textnormal{const} +
\int_\frac{r}{L}^1\!dx \frac{1}{x}\left(\frac{x}{2}\right)^{\left\vert n
\right\vert} \right\}
\end{equation}
where the negative sign refers to $n<0$.  For the case $n \neq 0$, the
infrared part converges and we are left with a function of the spatial
angle $\phi$ independent of $r$.  For the case $n=0$ we obtain a
logarithmic divergence in $r$.   Using the cut-off $r_c$ the precise form
for the $n=0$ contribution to the interaction functions is
\begin{equation}
\int\!\frac{d^2{\bf k} }{(2\pi)^2}\frac{e^{i{\bf k} \cdot{\bf r} }}{k^2} =
-\frac{1}{2\pi} \ln\left\vert \frac{r}{r_c} \right\vert
\end{equation}

Thus in the interaction terms only the zero-mode fourier component has any
mention of $r_c$.  To complete the renormalization group program, we must
write the partition function in a form containing only the renormalized
core size $r'_c = (1+dl)r_c$.  After we do this the partition function is
(up to an overall constant)
\begin{widetext}
\begin{eqnarray}
Z &=& \sum_{j=0}^{\infty} \sum_{k=0}^{\infty} y_N^{2j} y_M^{2k} (1+
dl)^{4k+4j}\frac{1}{j!^2} \frac{1}{k!^2} \int\!\frac{d^2{\bf
x}_1}{r^{\prime 2}_c}\cdots\frac{d^2{\bf x}_{2j}}{r^{\prime 2}_c}
\int\!\frac{d^2{\bf z}_1}{r^{\prime 2}_c}\cdots\frac{d^2{\bf z}_{2k}}{r^{\prime 2}_c} \times \nonumber\\
& & \exp(- \tilde{S'})\times \exp\left\{ - \frac{1}{2\pi} a'_0 \ln(1+dl)
\frac{1}{2} \sum_{\alpha\neq\beta} N_{\alpha} N_{\beta} - \frac{1}{2\pi}
\alpha^{\prime}_0 \ln(1+dl) \frac{1}{2} \sum_{\alpha\neq\beta} M_{\alpha}
M_{\beta} - \frac{i}{\pi} c'_0 \ln(1+dl) \frac{1}{2}
\sum_{\alpha\neq\beta} N_{\alpha} M_{\beta} \right\} \nonumber\\
\label{newZ}
\end{eqnarray}
\end{widetext} \noindent where $\tilde {S'}$ is the same as in equation
\ref{S}, but with $r_c$, $G_N$, $G_M$, and $\Gamma$ replaced by $r'_c$,
$G'_N$, $G'_M$, and $\Gamma'$, respectively.  The second exponential is
the correction from changing $r_c$ to $r'_c$ in the zero modes of the
interactions.

Now, by charge neutrality $\sum N_{\alpha}=\sum M_{\alpha} = 0$. Therefore
$\sum_{\alpha, \beta} M_{\alpha} N_{\beta} = 0$.  Also $\sum_{\alpha \neq
\beta} N_{\alpha} N_{\beta} = \left( \sum N_{\alpha} \right)^2 - \sum
N_{\alpha}^2 = -2j$ and similarly $\sum_{\alpha\neq\beta} M_{\alpha}
M_{\beta} = -2k$, so that the last exponential in equation \ref{newZ}
becomes $(1+dl)^{-2j\frac{a'_0}{4\pi}}
(1+dl)^{-2k\frac{\alpha^{\prime}_0}{4\pi}}$. With these contributions the
renormalization of $y_N$ and $y_M$ is
\begin{eqnarray}
y'_N &=& y_N + y_N (2-\frac{a_0}{4\pi}) dl \nonumber\\
y'_M &=& y_M + y_M (2-\frac{\alpha_0}{4\pi}) dl \label{ym}
\end{eqnarray}
to lowest order in the $y$.  At this point, the partition function is in
the same form as in equation \ref{Z} except that $r_c$ has been replaced
by $r'_c$, and the fourier coefficients and fugacities have changed
according to equations \ref{c} and \ref{ym}.  This completes the
renormalization group program and gives us the the differential
renormalization group flow equations \ref{flowend} stated in the text.

\end{document}